\documentclass[preprint,amsmath,amssymb,nofootinbib]{revtex4}

\newcommand{\be}{\begin{equation}}
\newcommand{\ee}{\end{equation}}
\newcommand{\ba}{\begin{eqnarray}}
\newcommand{\ea}{\end{eqnarray}}
\newcommand{\bn}{\begin{eqnarray*}}
\newcommand{\en}{\end{eqnarray*}}
\providecommand{\abs}[1]{\left\lvert#1\right\rvert}
\def\CF{{\cal F}}
\def\CE{{\cal E}}
\def\CB{{\cal B}}

\begin{document}

\title{SIM(2)-invariant Modifications of Electrodynamic Theory}

\author{Sangmo Chun} \email{haphysics@gmail.com}
\author{Choonkyu Lee} \email{cklee@phya.snu.ac.kr}
\author{Seung Jae Lee} \email{lsj8412@snu.ac.kr}

\affiliation{Department of Physics and Astronomy
and Center for Theoretical Physics,
Seoul National University, Seoul, 151-747, Korea}

\begin{abstract}
 In the Cohen-Glashow Very Special Relativity we exhibit possible
 modifications to the Maxwell theory and to the quantum electrodynamics
 Lagrangian in some generality, and discuss characteristic features
 depending on the modifications. Modified gauge transformations in SIM(2)-invariant
 theories are introduced and the related gauge fields, with
 two polarization states, can have nonzero mass.
 Also considered are SIM(2)-covariant modifications to
 the Proca-type field equations for a massive spin-1 particle.
\end{abstract}

\maketitle

\section{Introduction}

Special Relativity has been tested experimentally to a high degree of precision.
But recent years have seen renewed interest in possible tiny violations of
Lorentz symmetry
\cite{Kostelecky:1988zi,Carroll:1989vb,Coleman:1998ti,Carroll:2001ws,ArkaniHamed:2003uy,Colladay:1996iz},
as such effects are conceivable for instance in theories
that attempt to unify all known forces. In this regard, Cohen and Glashow \cite{VSR}
have recently made the ingenuous proposal that the laws of physics need not be invariant
under the full Lorentz group but rather under its SIM(2) subgroup, generated by
$T_1=K_x + J_y,\, T_2=K_y-J_x,\, J_z,$ and $K_z$
($\vec{J}$ and $\vec{K}$ are the generators of rotations and boosts, respectively).
This they referred to as Very Special Relativity(VSR).
In VSR space-time translational symmetry is retained so that the energy-momentum conservation,
and also the usual relativistic dispersion relation $E^2 = \vec{p}^{~2}+M^2$ for a particle of
mass M, may hold. SIM(2)-invariant, but Lorentz-violating, terms in the Lagrangian are
necessarily nonlocal and break discrete spacetime symmetries, including CP. Very recently there appeared
also
works devoted to continuous deformations of VSR \cite{Gibbons:2007iu} and a realization of VSR
via noncommutative deformation of Poincar\'{e} symmetry \cite{SheikhJabbari:2008nc}.

Some observable consequences of VSR have been studied in \cite{Cohen:2006ir,Fan:2006nd,Dunn:2006xk};
especially, in Ref. \cite{Cohen:2006ir}, a novel mechanism for neutrino masses without
introducing new particles has been given. This derives from the observation that a spin-$\tfrac{1}{2}$ particle
may satisfy the SIM(2)-covariant Dirac equation of the form
\be
 \label{eq:1}
 \left[ i\gamma^{\mu} \left(\partial_{\mu} + \frac{\lambda}{2}N_{\mu} \right)-m \right]\Psi(x)=0,
\ee
where $N_{\mu} \equiv \frac{n_{\mu}}{n \cdot \partial}$ with a chosen preferred null direction $n^{\mu}=(1,0,0,1)$.
If $\lambda \neq 0$, this equation is not Lorentz-covariant (and becomes nonlocal);
but, if one squares the SIM(2)-modified Dirac operator (and use $N \cdot N=0$, $N \cdot \partial =1$),
one obtains
\be
 \label{eq:2}
 \left[ \partial^{\mu}\partial_{\mu}+M^2 \right]\Psi(x)=0,~~~~(M^2=m^2+\lambda).
\ee
Therefore, even with $m=0$, the physical mass $M$ need not vanish if $\lambda \neq 0$.
(Here note that, with $m=0$, one may write (\ref{eq:1}) for a chirally projected field $\Psi(x)$).
A supersymmetric version of this theory was also considered in Refs.\cite{Cohen:2006sc,Lindstrom:2006xh}.

In this article we shall study possilbe SIM(2)-covariant modifications to the field equations of
a spin-1 particle.\footnote{We here continue to use the particle specification based on the Lorentz group
representation content, in the viewpoint that our Lorentz-violating, but SIM(2)-covariant, terms are to
be considered as perturbations to the usual Lorentz-covariant equations. For an alternative view as
regards this representation problem, see Ref.\cite{SheikhJabbari:2008nc}.}
This will be the first step to considering SIM(2)-covariant gauge theories.
Let us here recall the situation for the spin-1 field equation with the full Lorentz symmetry.
For a massive spin-1 particle it is given by the Proca equation
\be
 \label{eq:3}
 \partial_{\mu}F^{\mu \nu} + M^2 A^{\nu} = 0,~~~~(F^{\mu \nu}\equiv \partial^{\mu}A^{\nu}-\partial^{\nu}A^{\mu}).
\ee
Since $\partial_{\mu} \partial_{\nu} F^{\mu \nu} \equiv 0$, this is
is equivalent to the two equations,
\begin{subequations}\label{eq4}
\begin{eqnarray}
\label{eq:4a}
 \partial \cdot A=0,\\
\label{eq:4b}
 (\partial^2+M^2)A^{\mu}=0,
\end{eqnarray}
\end{subequations}
as long as $M^2\neq 0$. The Proca field thus has \emph{three} polarization states for given momentum.
Considering the $M=0$ limit of this system requires a care. As is well known \cite{Weinberg:1995mt},
the resulting system, i.e., the one described by Maxwell equations
(we have here included a current source $J^{\mu}$ also)
\be
 \label{eq:5}
 \partial_{\mu}F^{\mu \nu} = J^{\mu}
\ee
should be interpreted \emph{\`{a} la} gauge theory: any two gauge fields $A^{\mu}$ and $A'^{\mu}$
related by gauge transformation
\be
 \label{eq:6}
 A'^{\mu}(x) =  A^{\mu}(x) + \partial^{\mu} \Lambda(x)
\ee
do not refer to physically distinct states. As a result, for a massless spin-1 particle,
there are only \emph{two} physical polarizations. In VSR we will show that both the Proca-type and
Maxwell-type equations allow some nontrivial generalizations (with the latter as the appropriate singular
limit of the former), just as we have the modified Dirac equation for the spin-$\frac{1}{2}$ particle case.
Here it is possible to have a massive spin-1 particle having only two polarization states.
[The authors of Refs.\cite{Cohen:2006sc,Lindstrom:2006xh} also studied SIM(2)-covariant modifications
of (\ref{eq:5}) for their supersymmetric extension, but in a rather restricted form (by not
considering the possibility of modifying the gauge transformations (\ref{eq:6}) for instance)].

The rest of this letter is organized as follows. In the next section we consider the theory of a spin-1
particle satisfying SIM(2)-covariant Proca-type equations in some detail. The limiting case requiring a
gauge theory interpretation is also identified. Then, in section 3, we give the SIM(2)-invariant modifications
of the Maxwell-type theory, elaborate on the gauge symmetries in these theories, and
discuss gauge invariant interactions with matter fields in a manner consistent with VSR.
In section 4 we
summarize our findings and discuss possible phenomenological implications briefly.

\section{SIM(2)-invariant Proca-type theory}

For the SIM(2)-modified Proca equation we expect the SIM(2)-covarinat,
but nonlocal, vector operator $ N^{\mu} \equiv \frac{ n^{\mu}}{n \cdot \partial}$ to play a
key role.\footnote{
Using the lightcone coordinates
$x^{\pm} = \frac{1}{\sqrt{2}} (x^0 \pm x^3)$, we may write $N^{\mu}=\frac{n^\mu}{\sqrt{2}\partial_{+}}$
and thus specify its action on a function $f(x)$ by
\begin{equation*}
(N^{\mu} f)(x) \equiv \frac{n^{\mu}}{\sqrt{2}} \int d^4 y~G(x^{+}-y^{+}) \delta(x^1-y^1)\delta(x^2-y^2)\delta(x^{-} -y^{-}) f(y), \\
\end{equation*}
where $G(x^{+}-y^{+})=\theta(x^{+}-y^{+})-\frac{1}{2}~~(= - G(y^{+}-x^{+}))$. For these nonlocal operators, note that
$N \cdot N=0$, $N \cdot \partial = 1$, $[N^{\mu},N^{\nu}]=[N^{\mu},\partial^{\nu}]=0$, and
\begin{equation*}
 \int d^4 x~f(x)(N^{\mu}g)(x) = - \int d^4 x~ (N^{\mu}f)(x) g(x).
\end{equation*}
It would also be consistent to set $(N^{\mu}\Lambda)(x) \equiv 0$ if $\Lambda$ is a constant.
}
Here, in addition to the SIM(2) covariance, we will demand followings on our equation: (i) it should be
linear in $A^{\mu}(x)$, (ii) it reduces to the Proca equation (\ref{eq:3}) once all (small) parameters
in front of independent Lorentz-violating terms are set to zero, (iii) Lorentz-violating terms
in our equation may contain first or second derivatives (aside from $N^{\mu}$-factors) at most,
and (iv) solutions of this equation should also satisfy Eq.(\ref{eq:4b}) which is the mass shell
condition. Then, after executing some analysis to impose these requirements, we are led to
the $A_{\mu}$-equation of the general form
\be
 \label{eq:7}
 (\partial^2+M^2)A_{\nu}-(\partial_{\nu}+(g_1+g_3)N_{\nu})(\partial \cdot A + g_2 N \cdot A)
  + g_3(\partial^2+M^2)N_{\nu} N \cdot A = 0,
\ee
where $g_1$, $g_2$, $g_3$ (and also M) may assume any real values. This clearly fulfills the
requirements (i)-(iii). To check the consistency with our requirement (iv), note that
following equations
\begin{subequations}\label{eq:8}
\begin{eqnarray}
 \label{eq:8a}
 (M^2-g_1-g_2)(\partial \cdot A + g_2 N \cdot A) = 0, \\
 \label{eq:8b}
 (\partial^2+M^2)N \cdot A - (\partial \cdot A + g_2 N \cdot A)=0
\end{eqnarray}
\end{subequations}
are direct consequences of Eq.(\ref{eq:7}), being obtained if we
apply ($\partial^{\nu}+(g_2-g_3)N^{\nu}$) and $N^{\nu}$ from the
left, respectively. Now, if $M^2 \neq g_1+g_2$, we can combine
Eqs. (\ref{eq:8a}) and (\ref{eq:8b}) with Eq.(\ref{eq:7}) to
conclude that our SIM(2)-modified Proca equation Eq.(\ref{eq:7})
is equivalent to the following two equations
\begin{subequations}\label{eq9}
\begin{eqnarray}
 \label{eq:9a}
 \partial \cdot A + g_2 N \cdot A = 0, \\
 \label{eq:9b}
 (\partial^2+M^2) A_{\nu }=0.
\end{eqnarray}
\end{subequations}
Hence the correct mass shell condition is implied by our modified
Proca equation (\ref{eq:7}). When $M^2\neq g_1 + g_2$, on-shell
physical effects of Lorentz violating terms of Eq.(\ref{eq:7}) are
entirely contained  in Eq.(\ref{eq:9a}), which is the equation
determining the nature of three independent polarization modes;
i.e., for given momentum $p^{\mu}$, polarization vectors
$\epsilon^{\mu}$ must fulfill the condition
$q_{\mu}\epsilon^{\mu}=0$ where $q_{\mu} \equiv p_{\mu}-g_2
\frac{n_{\mu}}{n \cdot p}$. Explicitly, two of these polarization
directions may be chosen to be purely spatial, that is,
$\epsilon_{\pm}^{\mu}=(0,\vec{\epsilon}_{\pm})$ with
$\vec{\epsilon}_{\pm} \cdot  \vec{q}=0$, and then the third to be
$\epsilon_{L}^{\mu}=
\frac{1}{\sqrt{M^2-2g_2}}\frac{q^0}{\abs{\vec{~q~}}}\left(\frac{\abs{\vec{~q~}}^2}{q^0}
,\vec{q}~\right)$.

From Eqs.(\ref{eq:9a}) and (\ref{eq:9b}) we see that only the two parameters $M^2$ and $g_2$ in
Eq.(\ref{eq:7}) are physically relevant parameters (as long as the value of $g_1$ is not equal to
$M^2-g_2$). In such a situation an additional demand may be made on the form of our
equation (\ref{eq:7})---it should be derivable from a suitable action. This will obviously
be the case if we make the choice $g_3=g_2-g_1$, i.e., for the modified Proca equation
\be
 \label{eq:10}
 (\partial^2+M^2)A_{\nu}-(\partial_{\nu}+g_2 N_{\nu})(\partial \cdot A + g_2 N \cdot A)
  + (g_2-g_1)(\partial^2+M^2)N_{\nu} N \cdot A = 0,
\ee
which coincides with the stationary condition for the action
\ba
 \label{eq:11}
 S = \int d^4 x~ \left( - \frac{1}{2}(\partial^{\mu}A^{\nu})(\partial_{\mu}A_{\nu}) + \frac{1}{2}M^2 A^{\mu}A_{\mu} + \frac{1}{2}(\partial \cdot A + g_2 N \cdot A )^2 \right.  \nonumber \\
    \left. + \frac{1}{2}(g_2-g_1)\Bigl[(\partial^{\mu} N \cdot A)(\partial_{\mu} N \cdot A)-M^2(N\cdot A)^2 \Bigr] \right).
\ea
To facilitate our ensuing discussions, we may here define the tensor
\be
 \label{eq:12}
 \CF_{\mu \nu} \equiv (\partial_{\mu}+ g_1 N_{\mu})A_{\nu}-(\partial_{\nu}+ g_1 N_{\nu})A_{\mu} ~~(= - \CF_{\nu \mu})
\ee
Then the equation of motion (\ref{eq:10}) can be cast as
\be
 \label{eq:13}
 (\partial^{\mu}+g_2 N^{\mu})\CF_{\mu \nu} - (g_2-g_1)N_{\nu}N^{\mu}\partial^{\lambda}\CF_{\mu \lambda}
 +(M^2-g_1-g_2)\Bigl[ A_{\nu}+(g_2-g_1) N_{\nu}(N \cdot A)\Bigr]=0,
\ee
and the action (\ref{eq:11}) as
\ba
 \label{eq:14}
 S = &&\int d^4 x~ \left( - \frac{1}{4} \CF^{\mu \nu} \CF_{\mu \nu} + \frac{1}{2}(g_2-g_1) (N_{\lambda} \CF^{\lambda \nu})( N^{\mu}\CF_{\mu \nu}) \right.  \nonumber \\
    && ~~~~~~~~~~~~\left. + \frac{1}{2}(M^2-g_1-g_2)\Bigl[A^{\nu}A_{\nu}-(g_2-g_1)(N \cdot A)^2 \Bigr] \right).
\ea

Some comments are in order. If $g_1=M^2-g_2$, Eqs. (\ref{eq:9a}) and (\ref{eq:9b}) cannot be
deduced from Eq.(\ref{eq:10}) (or from Eq.(\ref{eq:7})): using the equivalent form in Eq.(\ref{eq:13}), this
is related to the gauge invariance of the system when the last term in the right-hand side of
Eq.(\ref{eq:13}) disappears. We study this singular limit in the next section. Another point is
that, upon making the change of field variables from $A_{\mu}(x)$ to $B_{\mu}(x)$ by
$A_{\mu}(x)=B_{\mu}(x)+(g_1+g_3-g_2)N_{\mu}N \cdot B(x)$, our equation (\ref{eq:7}) can be recast
into the form
\be
 \label{eq:15}
 (\partial^2+M^2)B_{\nu}-(\partial_{\nu}+g_2' N_{\nu})(\partial \cdot B + g_2' N \cdot B)
  + (g_2'-g_1')(\partial^2+M^2)N_{\nu} N \cdot B = 0
\ee
(here we set $g_1+g_3=g_2'$ and $g_2-g_3=g_1'$), which has the same appearance as Eq.(\ref{eq:10}).
Therefore, without loss of generality, we can take Eq.(\ref{eq:10}) or Eq.(\ref{eq:13}) as our SIM(2)-modified Proca
equation.

\section{SIM(2)-invariant abelian gauge theory}

Our equation (\ref{eq:13}) for $M^2=g_1+g_2~(>0)$, i.e.,
\be
 \label{eq:16}
 (\partial^{\mu}+g_2 N^{\mu})\CF_{\mu \nu} - (g_2-g_1)N_{\nu}N^{\mu}\partial^{\lambda}\CF_{\mu \lambda}=0
\ee
possesses gauge symmetry as the tensor $\CF_{\mu \nu}$,
given by Eq.(\ref{eq:12}), is invariant under the gauge transformation of the form
\be
 \label{eq:17}
 A_{\nu}(x) \longrightarrow   A'_{\nu}(x)  = A_{\nu}(x) + \partial_{\nu}\Lambda + g_1 (N_{\nu}\Lambda)(x)
\ee
where $\Lambda(x)$ can be an \emph{arbitrary} function of $x$. On the other hand,
based on Eq.(\ref{eq:16}) and the equation (which entails our definition (\ref{eq:12}) for $\CF_{\mu \nu}$)
\be
 \label{eq:18}
 \epsilon^{\mu \nu \lambda \delta} (\partial_{\nu}+g_1 N_{\nu}) \CF_{\lambda \delta} =0,
\ee
one can deduce that
\be
 \label{eq:19}
 \Bigl( \partial^2 + (g_1+g_2) \Bigr) \CF_{\mu \nu}=0,
\ee
i.e., gauge-invariant excitations here correspond to (in general) massive modes with $M^2=g_1+g_2$.
To prove Eq.(\ref{eq:19}), notice that (i) (by acting $N^{\nu}$ from the left) Eq.(\ref{eq:16}) implies
$N^{\nu} \partial^{\mu} \CF_{\mu \nu}=0$ and hence the equation
\be
 \label{eq:20}
 (\partial^{\mu} + g_2 N^{\mu})\CF_{\mu \nu} = 0
\ee
which contains $N^{\nu}\partial^{\mu}\CF_{\mu \nu}=0$, and (ii) we have, since Eq.(\ref{eq:18}) is
equivalent to the condition $(\partial+g_1 N)_{[\nu}\CF_{\lambda \delta ]}=0$,
\ba
 \label{eq:21}
 0&=& (\partial^{\mu}+g_2 N^{\mu})\Bigl[ (\partial_{\mu}+g_1 N_{\mu})\CF_{\lambda \delta} + (\partial_{\lambda}+g_1 N_{\lambda})\CF_{\delta \mu}+(\partial_{\delta}+g_1 N_{\delta})\CF_{\mu \lambda}\Bigr] \nonumber \\
  &=& \Bigl( \partial^2 + (g_1+g_2) \Bigr) \CF_{\lambda \delta}=0,
\ea
where we used Eq.(\ref{eq:20}). From these discussions it is also evident that our system given here
may be characterized entirely using gauge-invariant field strengths $\CF_{\mu \nu}$ only, i.e., by
the two Maxwell-like equations in Eqs. (\ref{eq:18}) and (\ref{eq:20}). (The special case of this model,
with $g_1=0$, was discussed in Refs.\cite{Cohen:2006sc, Lindstrom:2006xh}).

Disregarding the unphysical modes related to the gauge transformation (\ref{eq:17}),
there are now only \emph{two} independent polarization vectors for plane wave solutions
of Eq.(\ref{eq:16}). To exhibit their nature, let $\epsilon^{\mu}$ denote the polarization
vector of a solution $A_{\mu}(x)\propto \epsilon_{\mu}(p)e^{-i p \cdot x}$.
Then, from Eq.(\ref{eq:20}) (written as equations for $A_{\mu}$), $\epsilon_{\mu}(p)$ must
satisfy the condition
\be
 \label{eq:22}
 (p^2-g_1-g_2)\epsilon_{\mu} - \left( p_{\mu}-g_1 \frac{n_{\mu}}{n \cdot p}\right) \left( p\cdot \epsilon - g_2 \frac{n \cdot \epsilon}{n \cdot p}\right)=0.
\ee
If $p^2 \neq g_1 + g_2$, this shows that $\epsilon_{\mu} \propto \left( p_{\mu}-g_1 \frac{n_{\mu}}{n \cdot p}\right)$,
clearly a gauge excitation in view of Eq.(\ref{eq:17}). With $p^2=g_1+g_2$, on the other hand, Eq.(\ref{eq:22}) reduces
to the 4-dimensional orthogonality condition $q^{\mu} \epsilon_{\mu} =0$ where $q^{\mu} \equiv p^{\mu} - g_2 \frac{n_{\mu}}{n \cdot p}$
(with $q^2=g_1-g_2$). One vector satisfying this condition is $\epsilon_{\mu} \propto \left( p_{\mu}-g_1 \frac{n_{\mu}}{n \cdot p}\right)$,
a pure gauge again. Remaining two physical polarizations $\epsilon_{\mu}^{(i)}$ ($i=1,2$) may then be chosen such that
they satisfy the two conditions
\be
 \label{eq:23}
 \epsilon_{\mu}^{(i)} \left( p^{\mu}-g_1 \frac{n^{\mu}}{n \cdot p}\right) = 0, ~~~
 \epsilon_{\mu}^{(i)} \left( p^{\mu}-g_2 \frac{n^{\mu}}{n \cdot p}\right) = 0, ~~(i=1,2)
\ee
simultaneously. Note that, unless $g_1=g_2$, it will not be possible to take both vectors, i.e.,
$\epsilon_{\mu}^{(1)}(p)$ \emph{and} $\epsilon_{\mu}^{(2)}(p)$, to be purely spatial vectors ---
at least one of them has nonzero time component (for generic $\vec{p}$).

We may now introduce an (electric) source current $J_{\nu}$ and write the corresponding field equation
\be
 \label{eq:24}
 (\partial^{\mu}+g_2 N^{\mu})\CF_{\mu \nu} - (g_2-g_1)N_{\nu}N^{\mu}\partial^{\lambda}\CF_{\mu \lambda}=J_{\nu}
\ee
(for $\CF_{\mu\nu}$ still given by Eq.(\ref{eq:12}), i.e., with no modification on Eq.(\ref{eq:18})),
if $J_{\nu}$ satisfies the modified conservation law\footnote{ \label{fn:1}
From Eq.(\ref{eq:25}) one need not conclude that our theory does not allow a conserved current.
Actually, when Eq.(\ref{eq:25}) is true, another current $K^{\nu} \equiv J^{\nu}+g_1 N^{\nu} (N \cdot J)$
satisfies the usual conservation law $\partial_{\nu} K^{\nu}=0$. This is also
related to the fact that, by making a change of field variables analogous to that used in the last
paragraph of section 2, a different form of conservation law is obtained for the corresponding source current.
}
\be
 \label{eq:25}
 (\partial_{\nu}+g_1 N_{\nu})J^{\nu} = 0.
\ee
The condition (\ref{eq:25}) follows since the result after applying $(\partial^{\nu}+g_1 N^{\nu})$ on the
right hand side of Eq.(\ref{eq:24}) is identically zero. The SIM(2)-covariant field equation (\ref{eq:24})
can be derived by positing the action form
\be
 \label{eq:26}
 S = \int d^4 x~ \left( - \frac{1}{4} \CF^{\mu \nu} \CF_{\mu \nu} + \frac{1}{2}(g_2-g_1) (N_{\lambda} \CF^{\lambda \nu}) (N^{\mu}\CF_{\mu \nu}) - J^{\nu}A_{\nu}\right),
\ee
and, clearly, the condition (\ref{eq:26}) is what we need for the invariance of this action under the
local gauge transformation (\ref{eq:17}). We also remark that the field equation (\ref{eq:24}) can be presented by
the two Maxwell-like equations of the form
\begin{subequations}\label{eq27}
\begin{eqnarray}
 \label{eq:27a}
 (\partial^{\mu}+g_2 N^{\mu})\CF_{\mu \nu}&=&J_{\nu} - (g_2-g_1) N_{\nu} N \cdot J, \\
 \label{eq:27b}
 \epsilon^{\mu \nu \lambda \delta} (\partial_{\nu}+g_1 N_{\nu}) \CF_{\lambda \delta} &=&0.
\end{eqnarray}
\end{subequations}

The SIM(2)-invariant generalization of quantum electrodynamics, now involving some dynamical
current $J^{\mu}$, can also be given. For the sake of consistency with the condition (\ref{eq:25})
we must demand gauge invariance also on the matter part of the action.
Let us assume that the noninteracting Dirac field satisfies
Eq.(\ref{eq:1}) --- i.e., the action for the free Dirac field is
\be
 \label{eq:28}
 S_{Dirac} =  \int d^4 x~  \overline {\Psi}(x) \left[ i\gamma^{\mu} \left(\partial_{\mu} + \frac{\lambda}{2} \frac{n_{\mu}}{n \cdot \partial} \right)-m \right]\Psi(x).
\ee
Then, for the coupled system of this Dirac field and the above gauge field $A^{\mu}$,
the full action can be chosen as
\ba
 \label{eq:29}
 S  =  \int d^4 x~  \left[\overline {\Psi}(x) \left( i\gamma^{\mu} \left(D_{\mu} + \frac{\lambda}{2} \frac{n_{\mu}}{n \cdot D} \right)-m \right)\Psi(x) \right.\nonumber \\
 \left. - \frac{1}{4} \CF^{\mu \nu} \CF_{\mu \nu} + \frac{1}{2}(g_2-g_1) (N_{\lambda} \CF^{\lambda \nu}) (N^{\mu}\CF_{\mu \nu}) \right]
\ea
with the gauge covariant derivative $D_{\mu}$ given by
\be
 \label{eq:30}
 D_{\mu} = \partial_{\mu} + i e (A_{\mu}-g_1 N_{\mu} N \cdot A).
\ee
This action is gauge invariant, that is, invariant under the transformation (\ref{eq:17}),
$\Psi(x)\rightarrow \Psi '(x) = e^{-ie \Lambda(x)} \Psi(x)$, and
$\overline \Psi(x)\rightarrow \overline\Psi '(x) = e^{ie \Lambda(x)} \overline \Psi(x)$,
since the gauge transformations of $D_{\mu}$ and $\langle x |\frac{1}{n \cdot D} | y \rangle$,
i.e., $D'_{\mu}$ and $\langle x |\frac{1}{n \cdot D'} | y \rangle$ (with $A'_{\mu}$ given by Eq.(\ref{eq:17}))
are equal to $e^{-ie \Lambda(x)} D_{\mu} e^{ie \Lambda(x)}$ and
$e^{-ie \Lambda(x)}\langle x |\frac{1}{n \cdot D} | y \rangle e^{ie \Lambda(y)} $,
respectively.\footnote{
Note that, since $n \cdot N=0$, the last term in Eq.(\ref{eq:30}) does not enter $n \cdot D$.
Hence it is possible to write $\frac{1}{n \cdot D} =W  \frac{1}{n \cdot \partial} W^+$
by introducing the Wilson line \cite{Dunn:2006xk} $ W(x)=\exp \left( -ie \int_{-\infty}^{0} d s ~n^{\mu} [A_{\mu}(x+ n s)]\right)$.
}
In Eq.(\ref{eq:30}) we have a three-parameter (i.e., $\lambda, g_1$ and $g_2$) generalization
of usual quantum electrodynamics by demanding only the SIM(2) subgroup symmetry from
the full Lorentz group.

Note that our generalized quantum electrodynamics (\ref{eq:29}) is invariant under
global phase transformations on the fields $\Psi(x), \overline \Psi(x)$.
Therefore we have a (nonlocal) fermion current $K^{\mu}(x)$ which
satisfies the usual conservation law, $\partial_{\mu} K^{\mu}=0$.
With some calculations one can actually show that this fermion current
is related to our gauge-field source current $J^{\mu}\equiv \frac{\delta S}{\delta A_{\mu}(x)}$,
satisfying Eq.(\ref{eq:25}), by the equation (see the footnote \ref{fn:1})
$K^{\mu}=J^{\mu}+g_1 N^{\mu} (N \cdot J)$.

\section{Discussions}

To address an issue like Lorentz symmetry violations, it is important to have a definite
theoretical framework or model for the discussion. As for electrodynamics in particular,
some earlier developments in this regard include the Chern-Simons-like term addition \cite{Carroll:1989vb}
and the noncommutative-space generalization \cite{Carroll:2001ws}; these are models with
Lorentz-symmetry violation, but still gauge invariant. In this paper we formulated another
--- the SIM(2)-invariant electrodynamics --- according to the Cohen-Glashow VSR philosophy.

The SIM(2)-invariant electrodynamics features nonlocal terms with the directional dependence
due to the presence of a preferred null vector $n^{\mu}$.
One speculation will be that such terms might arise if there exist a certain, possibly cosmic,
medium of some unknown nature.
With no charged matter around, the field strengths satisfy the two-parameter (denoted $g_1$ and $g_2$)
extension of the usual Maxwell equations, given by Eqs. (\ref{eq:18}) and (\ref{eq:20}), and
gauge invariant excitations now acquire mass $M_{\gamma}=\sqrt{g_1+g_2}$.
(For photons we may thus demand $\sqrt{g_1 + g_2} < 10^{-18} eV$,
using the presently available experimental limit \cite{Goldhaber:2008xy}).
Further, the strict transversality for the associated wave solution no longer holds;
explicitly, if we consider plane waves with
$ E^{i}(x) \equiv \CF ^{ i 0 }(x) \propto  \CE^{i}(\vec{p~}) e^{i p \cdot x} $ and
$ B^{i}(x) \equiv \frac{1}{2} \epsilon^{ijk} \CF_{j k}(x) \propto  \CB^{i}(\vec{p~}) e^{i p \cdot x} $
(with $p^{0}=\sqrt{g_1+g_2+\vec{p~}^2}$), Eqs.(\ref{eq:18}) and (\ref{eq:20}) demand that
$\vec{\CE}(\vec{p})$ satisfy the condition
\be
 \label{eq:31}
 \vec{p} \cdot \vec{\CE} - g_2 \frac{\hat{z} \cdot \vec{\CE}}{n \cdot p}=0,
\ee
and $\vec{\CB}(\vec{p})$ be related to $\vec{\CE}(\vec{p})$ by
\be
 \label{eq:32}
 \vec{\CB} = \frac{1}{\left( p^0 - g_1 \frac{1}{n \cdot p}\right)}\left( \vec{p} \times \vec{\CE} - g_1 \frac{\hat{z} \times \vec{\CE}}{n \cdot p} \right)
\ee
(so that $\vec{p} \cdot \vec{\CB} - g_1 \frac{\hat{z} \cdot \vec{\CB}}{n \cdot p}=0$),
where $\hat z$ denotes the spatial direction picked by our preferred null vector $n^{\mu}$
and so $n \cdot p = p^0 - \hat z \cdot \vec{p}$. The SIM(2) (abelian) gauge field can
couple to the matter current which satisfies more general conservation law than usual,
our equation (\ref{eq:25}). Such example is provided by our three-parameter extension
of the usual quantum electrodynamics in Eq.(\ref{eq:29});
this can be a useful framework for the future discussion of Lorentz symmetry violations.

Finally, noting that our newly introduced terms typically involve factors like
$\frac{g_1}{n \cdot p}$ or $\frac{g_2}{n \cdot p}$ with $n \cdot p = \sqrt{\vec{p~}^2+M_{\gamma}^2}-p_z$
($M_{\gamma}=\sqrt{g_1+g_2}$) for on-shell `photons',
we will briefly explain in what sense these are \emph{small} compared to the usual ones.
Both $g_1$ and $g_2$ should
be very small here, but the situation may be somewhat different depending on
whether the ratio $r \equiv \frac{\max(\abs{g_1},\abs{g_2})}{M_{\gamma}^2}$ is very large (i.e., $\gg 1$) or not
(i.e., $ \lesssim 1$).
Here, when $\Delta$ denotes the experimental angular resolution for photons of energy $E=p^0$,
it should be reasonable to assume that $E \Delta \gg M_{\gamma}$ (as $E \Delta $ can be related to the momentum uncertainty).
We may then consider smeared values (indicated by $\langle ~~\rangle$ below) of $\vec{p}, \frac{g_1}{n \cdot p}$
and $\frac{g_2}{n \cdot p}$ over the angular resolution $\Delta$.
If $r \lesssim 1$, the given experimental conditions in fact guarantee that
\be
 \label{eq:33}
 \abs{\langle p^0\rangle},~ \abs{\langle p^3\rangle} ~\gg~ \abs{\langle \frac{g_1}{n \cdot p} \rangle},~ \abs{\langle \frac{g_2}{n \cdot p}\rangle}
\ee
and so all Lorentz-violating effects are indeed small. But, with $r\gg 1$ (i.e., $g_1 \approx -g_2$ and $M_{\gamma}^2 \ll \abs{~g_1}$),
the above experimental conditions are not sufficient to have the inequality (\ref{eq:33}).
In the latter case (which is perhaps phenomenologically more interesting), a short analysis involving
angle smearing shows following: if the angle $\theta = \measuredangle \vec{p}$ and $\hat{z}$ is not too small,
Eq. (\ref{eq:33}) holds as long as $E \Delta \gg \sqrt{\abs{~g_1}} ~~(\gg M_{\gamma})$;
at the angle near $\theta=0$, we need to assume additionally (for Eq. (\ref{eq:33})) that
$\ln (\frac{E^2 \Delta}{M^2_{\gamma}}) \ll \frac{E^2 \Delta^2}{2 \abs{g_1}} $.
There thus exists a wide range for phenomenological considerations.

\medskip\noindent\textbf{\slshape Acknowledgments}
We would like to thank
Ji-Haeng Huh, Sungjay Lee, Jong-Chul Park, Seo-Ree Park and Hyun Seok Yang for useful discussions.
This research was supported by Basic Science Research Program through the National Research Foundation
of Korea(NRF) funded by the Ministry of Education, Science and Technology(2009-0076297).



\end{document}